\documentclass[aps,preprint,showpacs,groupedaddress]{revtex4-1}
\usepackage{amssymb}
\usepackage{mathrsfs}
\usepackage{amsmath}
\usepackage{mathtools}
\usepackage{pstricks}
\usepackage{color}
\usepackage{graphicx}
\usepackage{amsthm}

\usepackage{physics}

\begin{document}


\title{\bf Unconventional minimal subtraction and Callan-Symanzik methods for Lorentz-violating}



\author{G. S. Silva}
\email{giofisicaufpi@gmail.com}
\author{P. R. S. Carvalho}
\email{prscarvalho@ufpi.edu.br}
\affiliation{\it Departamento de F\'\i sica, Universidade Federal do Piau\'\i, 64049-550, Teresina, PI, Brazil}





\begin{abstract}
We present an explicit analytical computation of the quantum corrections, at next-to-leading order, to the critical exponents. We employ for that the Unconventional minimal subtraction, recently proposed, and the Callan-Symanzik methods to probe the universality hypothesis by comparing the outcomes for the critical exponents evaluated in both methods and the ones calculated previously in massless theories renormalized at different renormalization schemes. Furthermore, the consistency of the former method is investigated for the first time in literature, to our knowledge. At the end, we compute the critical exponents at any loop level by an induction process and furnish the physical interpretation of the results.   
\end{abstract}


\maketitle


\section{Introduction}\label{Introduction}

\par Lorentz invariant (LI) field theories have been designed and successfully predicted results in very accurate agreement with experimental data in the past years \cite{ItzyksonZuber,PeskinSchroeder,Griffiths}. Despite their triumphs, Lorentz-violating (LV) versions of these theories have been proposed to search for possible violations of the referred symmetry. In the high energy particle physics scenario, the standard model of elementary particles and fields was extended to include LV effects. This extension culminated in the so called standard model extension \cite{PhysRevD.58.116002,PhysRevD.65.056006,PhysRevD.79.125019,PhysRevD.77.085006}. On the other hand, recent extensions of theories showing this symmetry breaking mechanism in low energy physics, particularly in phase transitions and critical phenomena, were proposed \cite{0295-5075-108-2-21001, doi:10.1142/S021797921550259, doi:10.1142/S0219887816500493}. In the latter situation, it was studied the universal behavior of systems undergoing a continuous phase transition. For that, specifically, they were applied field-theoretic renormalization group and $\epsilon$-expansion techniques for the computation, by an infinity induction process after an explicit finite next-to-leading order evaluation, of the all-loop quantum corrections to the critical exponents for massless LV O($N$) $\lambda\phi^{4}$ scalar field theories renormalized in the normalization conditions \cite{BrezinLeGuillouZinnJustin,Amit,ZinnJustin}, in the Bogoliubov-Parasyuk-Hepp-Zimmermann (BPHZ) \cite{BogoliubovParasyuk,Hepp,Zimmermann,Kleinert} and in the minimal subtraction \cite{Amit} methods, respectively. The critical exponents are universal quantities in the sense that they present identical results if they are evaluated in massless or massive theories renormalized at fixed or arbitrary external momenta, even if the theories are different at intermediate steps (if their renormalization constants, $\beta$-functions, anomalous dimensions, fixed points etc. are distinct). That is, in essence, the content of the universality hypothesis in the field-theoretic approach for the problem just mentioned. In this field-theoretic approach for the study of critical phenomena, massless and massive theories correspond to critical and noncritical theories, respectively, since the difference between an arbitrary temperature and the critical one is proportional to the mass of a quantum field. The fluctuations of this quantum field, when taken into account, give rise to the nontrivial quantum corrections to the classical mean field or Landau values for the critical exponents. The classical exponents can be easily obtained neglecting these fluctuations. As an example, for magnetic systems the order parameter is the magnetization. The magnetization is directly associated to the mean value of the fluctuating quantum field. In phenomenological terms, the critical exponents can be the same, thus confirming the universality hypothesis, even for completely distinct physical systems as a fluid and a ferromagnet. They do not depend on the microscopic details of the system, but just on the dimension $d$, $N$ and symmetry of some $N$-component order parameter and if the interactions between the degrees of freedom are of short- or long-range type. When the critical behavior of many systems is characterized by an identical set of critical exponents we say that they belong to the same universality class. The universality class inspected here, the O($N$) one, is a generalization of the particular models with short-range interactions: Ising ($N = 1$), XY ($N = 2$), Heisenberg($N = 3$), self-avoiding random walk ($N = 0$), spherical ($N \rightarrow\infty$) etc. \cite{Pelissetto2002549}. Many works considering the easily experimentally accessible physical parameters $d$ and $N$ were published \cite{PhysRevB.86.155112,PhysRevE.71.046112,PhysRevLett.110.141601,Butti2005527,PhysRevB.54.7177}. As the other concept from which the critical exponents also depend, the symmetry of the order parameter, is harder to probe than the earlier couple of them, we find just a few works on this subject \cite{PhysRevE.78.061124,Trugenberger2005509}. Thus the main aim in this work is to investigate the effect of a symmetry breaking mechanism, a LV one, on the critical exponents describing the critical behavior of systems belonging to the O($N$) universality class.

\par In this paper, we evaluate the all-loop quantum corrections to the critical exponents for the massive LV O($N$) $\lambda\phi^{4}$ scalar field theory by employing both Unconventional minimal subtraction \cite{:/content/aip/journal/jmp/54/9/10.1063/1.4819259} and the Callan-Symanzik methods \cite{PhysRevD.2.1541,symanzik1970,PhysRevD.8.4340}. In the latter method, the theory can be renormalized at the symmetry point in which the external momenta are set to null values, since a massive theory is not plagued by infrared divergences as in the massless case. A massless theory can not be defined at zero external momenta, i.e. the divergences are at the low momenta or infrared limit. The divergences to be treated here are in the opposite limit, in the ultraviolet or high momenta regime. Although a massive diagram is harder to evaluate than its massless counterpart, at a symmetry point in which its external momenta are fixed at vanishing values, this task is greatly simplified. Another simplification of this method is the drastic reduction in the number of diagrams to be evaluated, as we will show. While in the BPHZ method a finite large set of diagrams and counterterms, around of fifteen, had been used in the next-to-leading level LI computation \cite{Kleinert}, in the Callan-Symanzik method we have to evaluate a minimal set of just four of them. The same minimal set of diagrams is needed for attaining an identical loop order in the evaluation of the critical exponents in the former method, i. e. in the Unconventional minimal subtraction scheme \cite{:/content/aip/journal/jmp/54/9/10.1063/1.4819259}, which is a recently proposed and simpler alternative method to the BPHZ one. Thus, one of the aims of this work is to probe the consistency of this method in the LV scenario. The LV theory emerges when it is included into the LI Landau-Ginzburg Lagrangian density, the only LV O($N$) relevant operator (by power-counting analysis whose canonical dimension is less than or equal to four) $K_{\mu\nu}\partial^{\mu}\phi\partial^{\nu}\phi$. In general, $K_{\mu\nu}$ is a function of spacetime position and the present calculation would be very difficult in the general case. We then restrict the problem for the case where $K_{\mu\nu}$ is constant. In such situation, it is well known that the usual scalar field theory with Lorentz symmetry can be expressed in non-Minkowski coordinates. This was originally proved by V. Kosteleck\'{y} \cite{Phys.Rev.D.69.105009} and has been discussed in many papers since that one. We will use this fact for generalizing the next-to-leading critical exponents for their all loop orders counterparts. Higher power LV operators could be considered, but they would contribute with neglecting corrections to the critical behavior \cite{BF02819916,BF02742601,PhysRevD.7.2927}. They are called irrelevant operators and have canonical dimensions higher than four. Now the ultraviolet divergences of this LV theory are shown in the primitively one-particle irreducible ($1$PI) vertex parts $\Gamma^{(2)}$, $\Gamma^{(4)}$ and $\Gamma^{(2,1)}$. A renormalized theory is attained if these correlation functions are renormalized (the ultraviolet divergent higher $1$PI vertex parts can be written in terms of the primitively $1$PI correlation functions by an skeleton expansion, thus turning out them to be automatically renormalized). The critical exponents are then evaluated from the scaling properties of the renormalized primitively $1$PI vertex parts in a region near but not at the critical point. This is the case of a system which is not at the critical point but near it. Systems like that must be described by a massive or noncritical theory. The scaling properties give rise conditions to be satisfied between the critical exponents $\eta$ and $\nu$ and the field $\gamma_{\phi}$ and composite field $\overline{\gamma}_{\phi^{2}}$ anomalous dimensions computed at the fixed point, arising from the nontrivial solution of the $\beta$-function, respectively. The remaining four critical exponents can be computed from the four independent scaling relations among them.

\par This work begins with the renormalization of the theory and the subsequent computation of the critical exponents in the Unconventional minimal subtraction scheme in Sec. \ref{Unconventional minimal subtraction scheme}. In Sec. \ref{Callan-Symanzik method}, we achieve the same task by employing the Callan-Symanzik method. In Sec. \ref{All-loop quantum corrections to the critical exponents} we present an induction process leading to the evaluation of the all-loop quantum corrections to the critical exponents. At the end, we finalize the paper in Sec. \ref{Conclusions} with our conclusions.

\section{Unconventional minimal subtraction scheme}\label{Unconventional minimal subtraction scheme}

\par The theory considered here is described by the bare Lagrangian density, in the Euclidean spacetime suitable for statistical field theory computations, for the massive self-interacting LV O($N$) scalar field theory \cite{PhysRevD.84.065030,Carvalho2013850,Carvalho2014320} 
\begin{eqnarray}\label{bare Lagrangian density}
\mathscr{L} = \frac{1}{2}\partial^{\mu}\phi_{B}\partial_{\mu}\phi_{B} + \frac{1}{2}K_{\mu\nu}\partial^{\mu}\phi_{B}\partial^{\nu}\phi_{B} + \frac{1}{2}m_{B}^{2}\phi_{B}^{2}  +  \frac{\lambda_{B}}{4!}\phi_{B}^{4}.
\end{eqnarray}  
The unrenormalized parameters $\phi_{B}$, $m_{B}$ and $\lambda_{B}$ are the bare field, mass and coupling constant, respectively. The $N$-component quantum field interacts with itself through a quartic term of the form $\phi^{4} = (\phi_{1}^{2} + ... + \phi_{N}^{2})^{2}$. Although the Lagrangian density is invariant under transformations of elements of the O($N$) Euclidean orthogonal group, it is not under the Lorentz group. The term which breaks the latter symmetry is the second one in Eq. \ref{bare Lagrangian density}, where the symmetric constant coefficients $K_{\mu\nu}$ are responsible for a slight violation of the corresponding symmetry for $|K_{\mu\nu}|\ll 1$. Now the bare primitively $1$PI vertex functions can be expanded up to next-to-leading order by applying the following Feynman rules for the free propagator and vertices
\begin{eqnarray}
\parbox{14mm}{\includegraphics[scale=1.0]{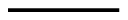}} = ~(q^{2} + K_{\mu\nu}q^{\mu}q^{\nu} + m_{B}^{2})^{-1},
\end{eqnarray}
\begin{eqnarray}
\parbox{12mm}{\includegraphics[scale=.1]{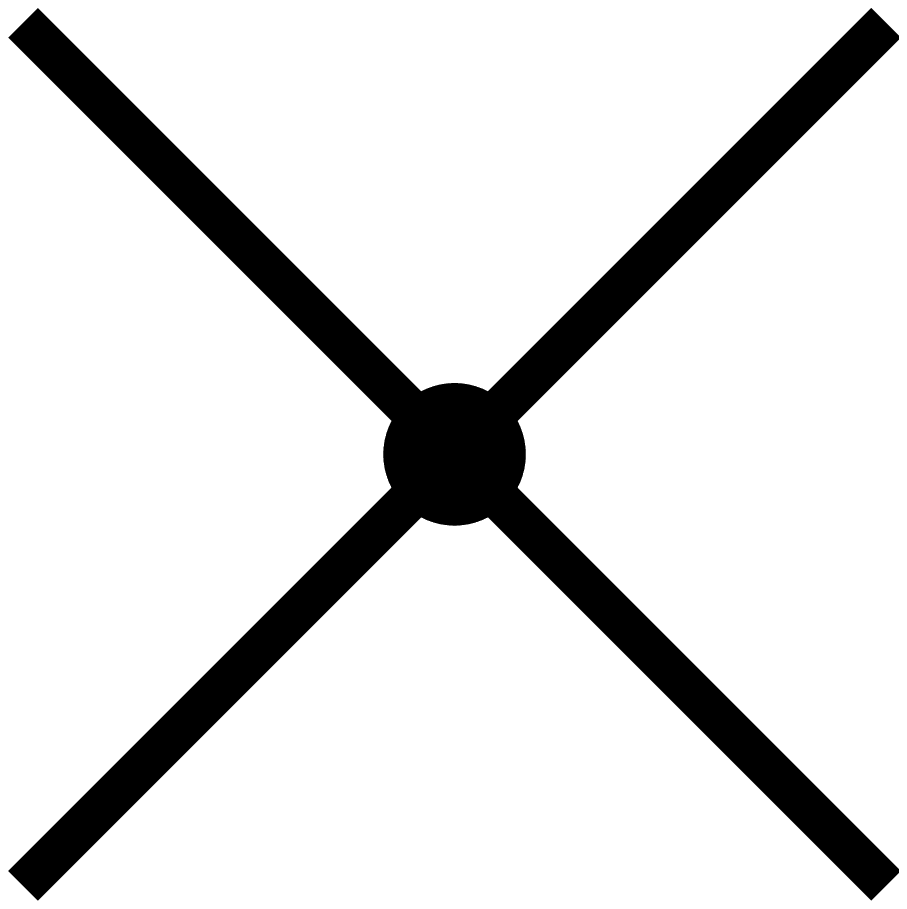}} = ~-\lambda_{B},
\end{eqnarray}
\begin{eqnarray}
\parbox{12mm}{\includegraphics[scale=1.0]{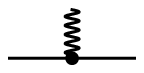}} ~~= ~~1.
\end{eqnarray}
The expansion is composed of many terms. It can be reduced to a small number of them, since all diagrams containing tadpole insertions 
\begin{eqnarray}
\parbox{12mm}{\includegraphics[scale=1.0]{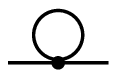}} \parbox{12mm}{\includegraphics[scale=1.0]{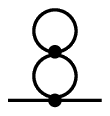}} \parbox{10mm}{\includegraphics[scale=1.0]{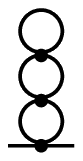}} \parbox{14mm}{\includegraphics[scale=1.0]{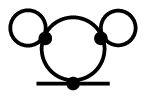}} \parbox{14mm}{\includegraphics[scale=1.0]{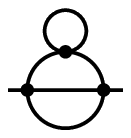}} \parbox{12mm}{\includegraphics[scale=1.0]{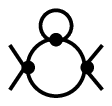}} \parbox{12mm}{\includegraphics[scale=1.0]{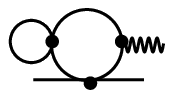}}
\end{eqnarray}
and the one which is independent of external momenta 
\begin{eqnarray}
\parbox{12mm}{\includegraphics[scale=1.0]{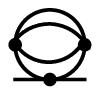}} 
\end{eqnarray}
can be eliminated. In fact, it was shown that this can achieved if we substitute the bare mass $m_{B,tree-level}$ in Eq. \ref{bare Lagrangian density} initially at tree-level for its three-loop counterpart $m_{B,three-loop}$ \cite{Amit, :/content/aip/journal/jmp/54/9/10.1063/1.4819259}. Thus just naming $m_{B,three-loop} \rightarrow m_{B}$ from now on we have for $N = 1$

\begin{eqnarray}
&&\Gamma_{B}^{(2)}(P,m_{B},\lambda_{B})  =  \quad \parbox{12mm}{\includegraphics[scale=1.0]{fig9.eps}}^{-1} \quad - \quad  \quad
\frac{\lambda_{B}^2}{6}\left(
\parbox{12mm}{\includegraphics[scale=1.0]{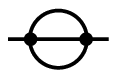}}  \quad - \quad
\parbox{12mm}{\includegraphics[scale=1.0]{fig6.eps}}\bigg|_{P=0}\right)
 \nonumber \\ && \quad  + \quad  \frac{\lambda_{B}^{3}}{4}\left(
\parbox{12mm}{\includegraphics[scale=.9]{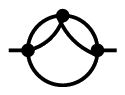}}  \quad - \quad
\parbox{12mm}{\includegraphics[scale=.9]{fig7.eps}}\bigg|_{P=0}\right), \label{14}
\end{eqnarray}

\begin{eqnarray}
&&\Gamma_{B}^{(4)}(P_{i},m_{B},\lambda_{B}) =  \lambda_{B} - \frac{\lambda_{B}^{2}}{2}
 \left(\parbox{10mm}{\includegraphics[scale=1.0]{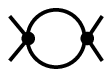}} + 2\hspace{2mm}perm.\right) \nonumber\\
&& + \frac{\lambda_{B}^{3}}{4}
         \left(\parbox{16mm}{\includegraphics[scale=1.0]{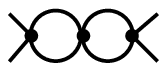}} + 2\hspace{2mm}perm.\right) + \frac{\lambda_{B}^{3}}{2}
   \left(\parbox{16mm}{\includegraphics[scale=.9]{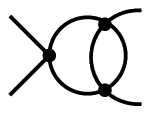}} + 5\hspace{2mm}perm. \right), \label{20}
\end{eqnarray}

\begin{eqnarray}
&& \Gamma_{B}^{(2,1)}(P_{1},P_{2},Q_{3},m_{B},\lambda_{B}) = 1 - 
\frac{\lambda_{B}}{2} \parbox{13mm}{\includegraphics[scale=1.0]{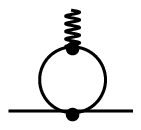}}  + \frac{\lambda_{B}^{2}}{4}\parbox{11mm}{\includegraphics[scale=1.0]{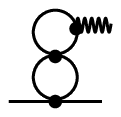}}  + \frac{\lambda_{B}^{2}}{2}\parbox{13mm}{\includegraphics[scale=.9]{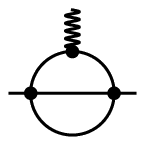}}\quad , \label{26}
\end{eqnarray}   
where $Q = -(P_{1} + P_{2})$. We can define a relation between the dimensional and the dimensionless renormalized coupling constants $\lambda$ and $u$ as $\lambda = u m^{\epsilon}$, where $m$, at the loop level considered here, is used as an arbitrary momentum scale and $\epsilon = 4 - d$ is the small parameter for which we will expand the dimensionally divergent diagrams around $d = 4$. An analog relation between the corresponding bare quantities $\lambda_{B}$ and $u_{0}$ can be also defined as $\lambda_{B} = u_{0}m^{\epsilon}$. Now the Eqs. above can be written, for general $N$, as 
\begin{eqnarray}\label{gtfrdrdes}
\Gamma^{(2)}_{B}(P, u_{0}, m_{B}) = P^{2}( 1 - B_{2}u_{0}^{2} + B_{3}u_{0}^{3}),
\end{eqnarray}
\begin{eqnarray}
\Gamma^{(4)}_{B}(P_{i}, u_{0}, m_{B}) = m_{B}^{\epsilon}u_{0}[ 1 - A_{1}u_{0} + (A_{2}^{(1)} + A_{2}^{(2)})u_{0}^{2}],
\end{eqnarray}
\begin{eqnarray}\label{gtfrdesuuji}
\Gamma^{(2,1)}_{B}(P_{1}, P_{2}, Q_{3}, u_{0}, m_{B}) = 1 - C_{1}u_{0} + (C_{2}^{(1)} + C_{2}^{(2)})u_{0}^{2},
\end{eqnarray}
where
\begin{eqnarray}
A_{1} = \frac{(N + 8)}{18}\left[ \parbox{10mm}{\includegraphics[scale=1.0]{fig10.eps}} + 2 \hspace{2mm} \text{perm.} \right]
\end{eqnarray}
\begin{eqnarray}
A_{2}^{(1)} = \frac{(N^{2} + 6N + 20)}{108}\left[ \parbox{10mm}{\includegraphics[scale=1.0]{fig11.eps}}\quad\quad  + 2 \hspace{2mm} \text{perm.} \right],
\end{eqnarray}
\begin{eqnarray}
A_{2}^{(2)} = \frac{(5N + 22)}{54}\left[ \parbox{10mm}{\includegraphics[scale=1.0]{fig21.eps}}\quad+ 5 \hspace{2mm} \text{perm.} \right],
\end{eqnarray}
\begin{eqnarray}
B_{2} = \frac{(N + 2)}{18}\hspace{1mm}\left(\parbox{12mm}{\includegraphics[scale=1.0]{fig6.eps}} \quad - \quad
\parbox{12mm}{\includegraphics[scale=1.0]{fig6.eps}}\bigg|_{P=0}\right),
\end{eqnarray}
\begin{eqnarray}
B_{3} = \frac{(N + 2)(N + 8)}{108}\hspace{1mm}\left(\parbox{12mm}{\includegraphics[scale=1.0]{fig7.eps}} \quad - \quad
\parbox{12mm}{\includegraphics[scale=1.0]{fig7.eps}}\bigg|_{P=0}\right),
\end{eqnarray}
\begin{eqnarray}
C_{1} = \frac{(N + 2)}{6}\hspace{1mm}\parbox{10mm}{\includegraphics[scale=1.0]{fig14.eps}}\hspace{4mm}, 
\end{eqnarray}
\begin{eqnarray}
C_{2}^{(1)} = \frac{(N + 2)^{2}}{36}\hspace{1mm}\parbox{10mm}{\includegraphics[scale=1.0]{fig16.eps}}\hspace{2mm},
\end{eqnarray}
\begin{eqnarray}
C_{2}^{(2)} = \frac{(N + 2)}{6}\hspace{1mm}\parbox{10mm}{\includegraphics[scale=1.0]{fig17.eps}}\hspace{5mm}.
\end{eqnarray}
We do not need to evaluate all the diagrams above. In fact, some of them are not independent. This is another fact that turns this method simpler than the BPHZ one, where we have to evaluate more than fifteen diagrams for performing an identical achievement. As the calculation of the loop contributions to the $1$PI vertex parts leads to momentum integration involving just their internal bubbles and not their external legs, all what matters are their internal bubbles contents. Thus, without take into account the O($N$) symmetry factors, we have that \parbox{7mm}{\includegraphics[scale=0.5]{fig10.eps}}$\propto$\hspace{1mm}\parbox{7mm}{\includegraphics[scale=0.5]{fig14.eps}}, \parbox{9mm}{\includegraphics[scale=0.5]{fig21.eps}}$\propto$\hspace{1mm}\parbox{8mm}{\includegraphics[scale=0.5]{fig17.eps}}, \parbox{7mm}{\includegraphics[scale=0.5]{fig11.eps}}\hspace{2mm}$\propto$\hspace{2mm}\parbox{7mm}{\includegraphics[scale=0.5]{fig16.eps}}\hspace{1mm}$\propto$\hspace{2mm}(\parbox{5mm}{\includegraphics[scale=0.5]{fig10.eps}})$^{2}$. Finally, the only diagrams to be considered are the $\parbox{6mm}{\includegraphics[scale=0.5]{fig10.eps}}$, $\parbox{8mm}{\includegraphics[scale=0.7]{fig6.eps}}$, $\parbox{7mm}{\includegraphics[scale=0.6]{fig7.eps}}$, $\parbox{7mm}{\includegraphics[scale=0.5]{fig21.eps}}$ ones, 
\begin{eqnarray}
&&\parbox{12mm}{\includegraphics[scale=1.0]{fig10.eps}} = \int \frac{d^{d}q}{(2\pi)^{d}}\frac{1}{q^{2} + K_{\mu\nu}q^{\mu}q^{\nu} + m_{B}^{2}} \frac{1}{(q + P)^{2} + K_{\mu\nu}(q + P)^{\mu}(q + P)^{\nu} + m_{B}^{2}}, \nonumber \\
\end{eqnarray}   
\begin{eqnarray}
&&\parbox{12mm}{\includegraphics[scale=1.0]{fig6.eps}} = \int \frac{d^{d}q_{1}}{(2\pi)^{d}}\frac{d^{d}q_{2}}{(2\pi)^{d}}\frac{1}{q_{1}^2 + K_{\mu\nu}q_{1}^{\mu}q_{1}^{\nu} + m_{B}^{2}} \frac{1}{q_{2}^2 + K_{\mu\nu}q_{2}^{\mu}q_{2}^{\nu} + m_{B}^{2}}  \nonumber \\&&  \frac{1}{(q_{1} + q_{2} + P)^2 + K_{\mu\nu}(q_{1} + q_{2} + P)^{\mu}(q_{1} + q_{2} + P)^{\nu} + m_{B}^{2}}, 
\end{eqnarray}   
\begin{eqnarray}
&&\parbox{12mm}{\includegraphics[scale=1.0]{fig7.eps}} = \int \frac{d^{d}q_{1}}{(2\pi)^{d}}\frac{d^{d}q_{2}}{(2\pi)^{d}}\frac{d^{d}q_{3}}{(2\pi)^{d}}\nonumber \\ &&\times  \frac{1}{q_{1}^2 + K_{\mu\nu}q_{1}^{\mu}q_{1}^{\nu} + m_{B}^{2}} \frac{1}{q_{2}^2 + K_{\mu\nu}q_{2}^{\mu}q_{2}^{\nu} + m_{B}^{2}}\frac{1}{q_{3}^2 + K_{\mu\nu}q_{3}^{\mu}q_{3}^{\nu} + m_{B}^{2}} \nonumber \\ &&\times\frac{1}{(q_{1} + q_{2} + P)^2 + K_{\mu\nu}(q_{1} + q_{2} + P)^{\mu}(q_{1} + q_{2} + P)^{\nu} + m_{B}^{2}} \nonumber \\ &&\times\frac{1}{(q_{1} + q_{3} + P)^2 + K_{\mu\nu}(q_{1} + q_{3} + P)^{\mu}(q_{1} + q_{3} + P)^{\nu} + m_{B}^{2}}, 
\end{eqnarray} 
\begin{eqnarray}
&&\parbox{14mm}{\includegraphics[scale=1.0]{fig21.eps}} = \int \frac{d^{d}q_{1}}{(2\pi)^{d}}\frac{d^{d}q_{2}}{(2\pi)^{d}}\frac{1}{q_{1}^2 + K_{\mu\nu}q_{1}^{\mu}q_{1}^{\nu} + m_{B}^{2}} \nonumber \\ && \times\frac{1}{(P - q_{1})^{2} + K_{\mu\nu}(P - q_{1})^{\mu}(P - q_{1})^{\nu} + m_{B}^{2}} \frac{1}{q_{2}^2 + K_{\mu\nu}q_{2}^{\mu}q_{2}^{\nu} + m_{B}^{2}} \nonumber \\ && \times\frac{1}{(q_{1} - q_{2} + Q_{3})^2 + K_{\mu\nu}(q_{1} - q_{2} + Q_{3})^{\mu}(q_{1} - q_{2} + Q_{3})^{\nu} + m_{B}^{2}}. 
\end{eqnarray}     
whose values are shown in \ref{Feynman integrals}. The $1$PI vertex parts are multiplicatively renormalized
\begin{eqnarray}\label{uhygtfrd}
\Gamma_{R}^{(n, l)}(P_{i}, Q_{j}, u, m) = Z_{\phi}^{n/2}Z_{\phi^{2}}^{l}\Gamma_{B}^{(n, l)}(P_{i}, Q_{j}, \lambda_{B}, m_{B})
\end{eqnarray}
where the poles are minimally eliminated, thus being absorbed in the renormalization constants for the field $Z_{\phi}$ and composite field $Z_{\phi^{2}}$, respectively. Writing the Laurent expansion 
\begin{eqnarray}\label{suhufjifjvf}
u_{0} = u\left( 1 + \sum_{i=1}^{\infty} a_{i}(\epsilon)u^{i}\right),
\end{eqnarray}
\begin{eqnarray}
Z_{\phi} = 1 + \sum_{i=1}^{\infty} b_{i}(\epsilon)u^{i},
\end{eqnarray}
\begin{eqnarray}\label{iaifkdvkvkck}
\overline{Z}_{\phi^{2}} = 1 + \sum_{i=1}^{\infty} c_{i}(\epsilon)u^{i},
\end{eqnarray}
where the renormalization constant $\overline{Z}_{\phi^{2}} \equiv Z_{\phi}Z_{\phi^{2}}$ is used instead of $Z_{\phi^{2}}$, for convenience, the renormalized $1$PI vertex parts  satisfy to the Callan-Symanzik equation
\begin{eqnarray}\label{udhuhsufdh}
&& \left( m\frac{\partial}{\partial m} + \beta\frac{\partial}{\partial u} - \frac{1}{2}n\gamma_{\phi} + l\gamma_{\phi^{2}} \right)\Gamma_{R}^{(n, l)}(P_{i}, Q_{j}, u, m) = \nonumber\\ && m^{2}(2 - \gamma_{\phi})\Gamma_{R}^{(n, l + 1)}(P_{i}, Q_{j}, 0, u, m)
\end{eqnarray}
where 
\begin{eqnarray}\label{kjjffxdzs}
\beta(u) = m\frac{\partial u}{\partial m} = -\epsilon\left(\frac{\partial\ln u_{0}}{\partial u}\right)^{-1},
\end{eqnarray}
\begin{eqnarray}\label{koiuhygtf}
\gamma_{\phi}(u) = \beta(u)\frac{\partial\ln Z_{\phi}}{\partial u},
\end{eqnarray}
\begin{eqnarray}\label{koiuhygtfddddd}
\gamma_{\phi^{2}}(u) = -\beta(u)\frac{\partial\ln Z_{\phi^{2}}}{\partial u}.
\end{eqnarray}
We use the function
\begin{eqnarray}\label{udgygeykoiuhygtf}
\overline{\gamma}_{\phi^{2}}(u) = -\beta(u)\frac{\partial\ln \overline{Z}_{\phi^{2}}}{\partial u} \equiv \gamma_{\phi^{2}}(u) - \gamma_{\phi}(u)
\end{eqnarray}
instead of $\gamma_{\phi^{2}}(u)$, for convenience reasons. While in the right-hand side (rhs) of Eq. \ref{udhuhsufdh}, the $1$PI vertex part has $l + 1$ composite field insertions, the one in the left-hand side (lhs) possess $l$ such insertions. We know that an extra composite field insertion represents one additional power of the propagator in the respective $1$PI vertex part. Thus in the ultraviolet limit, i. e., when the external momenta $P_{i}/m \rightarrow \infty$, the rhs can be neglected in comparison with the lhs, order by order in perturbation theory. This result can be seen as an application of the Weinberg's theorem \cite{PhysRev.118.838}. Then, in the ultraviolet limit, the $1$PI vertex parts satisfy the renormalization group equation and we can apply the entire theory of scaling for these functions and evaluate, perturbativelly, the $\beta$-function and anomalous dimensions as
\begin{eqnarray}
\beta(u) = -\epsilon u[ 1 - a_{1}u + 2(a_{1}^{2} - a_{2})u^{2} ],
\end{eqnarray}
\begin{eqnarray}
\gamma_{\phi}(u) = -\epsilon u[ 2b_{2}u + (3b_{3} - 2b_{2}a_{1})u^{2} ],
\end{eqnarray}
\begin{eqnarray}
\overline{\gamma}_{\phi^{2}}(u) = \epsilon u[ c_{1} + (2c_{2} - c_{1}^{2} - 2a_{1}c_{1})u ].
\end{eqnarray}
The constant coefficients $a_{1}$, $\cdots$, $c_{2}$ depend on the Feynman diagrams just mentioned \cite{:/content/aip/journal/jmp/54/9/10.1063/1.4819259}. Now absorbing $\hat{S}$ in a redefinition of the coupling constant and using the Feynman diagrams computed in \ref{Feynman integrals}, we find 
\begin{eqnarray}\label{uahuahuahu}
\beta(u) = u\left( -\epsilon + \frac{N + 8}{6}\Pi u - \frac{3N + 14}{12}\Pi^{2}u^{2}\right),
\end{eqnarray}
\begin{eqnarray}\label{uahuahuahuwesc}
\gamma_{\phi}(u) = \frac{N + 2}{72}\left( \Pi^{2}u^{2} - \frac{N + 8}{24}\Pi^{3}u^{3} \right),
\end{eqnarray}
\begin{eqnarray}\label{uahuahuahuaa}
\overline{\gamma}_{\phi^{2}}(u) = \frac{N + 2}{6}\left( \Pi u - \frac{1}{2}\Pi^{2}u^{2} \right).
\end{eqnarray}
where the LV $\Pi$ factor is given by
\begin{eqnarray}
\Pi = 1 - \frac{1}{2}K_{\mu\nu}\delta^{\mu\nu} + \frac{1}{8}K_{\mu\nu}K_{\rho\sigma}\delta^{\{\mu\nu}\delta^{\rho\sigma\}} + ...,
\end{eqnarray}
$\delta^{\{\mu\nu}\delta^{\rho\sigma\}} \equiv \delta^{\mu\nu}\delta^{\rho\sigma} + \delta^{\mu\rho}\delta^{\nu\sigma} + \delta^{\mu\sigma}\delta^{\nu\rho}$ and $\delta^{\mu\nu}$ is the Kronecker delta symbol. As in this method the renormalization is elegantly performed at arbitrary external momenta, all the momentum-dependent integrals $L(P, m_{B}^{2})$, $L^{\mu\nu}(P, m_{B}^{2})$, $L_{3}(P, m_{B}^{2})$, $L_{3}^{\mu\nu}(P, m_{B}^{2})$, $L^{\mu\nu\rho\sigma}(P, m_{B}^{2})$, $\tilde{i}(P, m_{B}^{2})$ and $\tilde{i}^{\mu\nu}(P, m_{B}^{2})$ have disappeared. This produces a LV functional dependence of the $\beta$-function, field and composite field anomalous dimensions on $K$, only through $\Pi$, in a power-law form. We will use this fact in further discussions. While the cancelling of the integrals $L_{3}(P, m_{B}^{2})$, $L_{3}^{\mu\nu}(P, m_{B}^{2})$ and $\tilde{i}^{\mu\nu}(P, m_{B}^{2})$ is responsible for the field renormalization at the loop level inspected here, shown in Eq. (\ref{uahuahuahuwesc}), since the renormalization of this parameter comes from the terms proportional to $P^{2}$ in the diagrams $\parbox{10mm}{\includegraphics[scale=.8]{fig6.eps}}$ and $\parbox{10mm}{\includegraphics[scale=.7]{fig7.eps}}$, we are yet left with a spurious divergence. In fact, it comes from the terms proportional to $m^{2}$ in the diagrams $\parbox{10mm}{\includegraphics[scale=.8]{fig6.eps}}$ and $\parbox{10mm}{\includegraphics[scale=.7]{fig7.eps}}$. These terms turn out the respective $1$PI vertex part a divergent one    
\begin{eqnarray}
&& \Gamma_{R}^{(2)}(P, u, m) = P^{2} + m^{2}\left\{1 + \frac{(N+2)}{24}[\tilde{I}(P, m_{B}^{2}) + K_{\mu\nu}\tilde{i}^{\mu\nu}(P, m_{B}^{2})]u^{2} - \right.  \nonumber \\  &&\left. \frac{(N+2)(N+8)}{108\epsilon}[\tilde{I}(P, m_{B}^{2}) + K_{\mu\nu}\tilde{i}^{\mu\nu}(P, m_{B}^{2})]u^{3} \right\},
\end{eqnarray}
where
\begin{eqnarray}
&&\tilde{I}(P, m_{B}^{2}) = \int_{0}^{1} dx \int_{0}^{1}dy lny \frac{d}{dy}\left\{(1-y)ln\left[\frac{y(1-y)\frac{P^{2}}{m_{B}^{2}} + 1-y + \frac{y}{x(1-x)}}{1-y + \frac{y}{x(1-x)}} \right]\right\}. \nonumber \\
\end{eqnarray}
This is the effect of reducing the number of diagrams just from the beginning, thus redefining the initial bare mass at tree-level to its three-loop order counterpart. To ensure that the method presented here subtracts the poles minimally, as claimed at the very beginning, we have to redefine the two-point function as   
\begin{eqnarray}
&& \tilde{\Gamma}_{R}^{(2)}(P, u, m) = \Gamma_{R}^{(2)}(P, u, m) + \nonumber \\  && m^{2}\left\{ \frac{(N+2)(N+8)}{108\epsilon}[\tilde{I}(P, m_{B}^{2}) + K_{\mu\nu}\tilde{i}^{\mu\nu}(P, m_{B}^{2})]u^{3} \right\}.
\end{eqnarray}
Thus, we make an intimate connection between the Unconventional minimal subtraction scheme and the conventional one applied for the massless theory \cite{Amit}, once the terms proportional to $m^{2}$ are null in the latter situation. One obvious but important check of this redefinition is that it satisfies the normalization condition used in the Sec. \ref{Callan-Symanzik method}
\begin{eqnarray}
&& \tilde{\Gamma}_{R}^{(2)}(0, u, m) = \Gamma_{R}^{(2)}(0, u, m) = m^{2}.
\end{eqnarray}

\ Now, for computing the LV loop quantum corrections to the critical exponents, we need to evaluate the so called nontrivial fixed point. It can be obtained as the nontrivial solution to the equation $\beta(u^{*}) = 0$. The trivial or Gaussian one gives only the classical or Landau mean field approximation values to the critical exponents, without taking into account their radiative quantum contributions. The nontrivial solution is given by   
\begin{eqnarray}\label{yagyaguhd}
u^{*} = \frac{6\epsilon}{(N + 8)\Pi}\left\{ 1 + \epsilon\left[ \frac{3(3N + 14)}{(N + 8)^{2}} \right]\right\}.
\end{eqnarray}
The definitions $\eta\equiv\gamma_{\phi}(u^{*})$ and $\nu^{-1}\equiv 2 - \eta - \overline{\gamma}_{\phi^{2}}(u^{*})$ can be applied to obtain, to next-to-leading order, the two respective critical exponents
\begin{eqnarray}\label{eta}
\eta = \frac{(N + 2)\epsilon^{2}}{2(N + 8)^{2}}\left\{ 1 + \epsilon\left[ \frac{6(3N + 14)}{(N + 8)^{2}} -\frac{1}{4} \right]\right\}\equiv\eta^{(0)},
\end{eqnarray}
\begin{eqnarray}\label{nu}
&&\nu = \frac{1}{2} + \frac{(N + 2)\epsilon}{4(N + 8)} +  \frac{(N + 2)(N^{2} + 23N + 60)\epsilon^{2}}{8(N + 8)^{3}}\equiv\nu^{(0)},
\end{eqnarray}
where $\eta^{(0)}$ and $\nu^{(0)}$ are their corresponding LI counterparts \cite{Wilson197475}. As there are six critical exponents and four scaling relations among them, there are only two of them which are independent. Thus the two ones above are enough for evaluating the four remaining ones. In next Sec. we will attain the same task but now in a distinct renormalization method and will compare the results.     
  
\section{Callan-Symanzik method}\label{Callan-Symanzik method}

In the Callan-Symanzik method, the diagrams $\parbox{10mm}{\includegraphics[scale=.8]{fig6.eps}}\bigg|_{P=0}$ and $\parbox{10mm}{\includegraphics[scale=.7]{fig7.eps}}\bigg|_{P=0}$, while evaluated at fixed vanishing external momenta, do not depend on the external momenta of the respective diagrams. This implies that they will not contribute to the subsequent computations, since the method approached here deals with the derivative of them with respect to their external momenta. The derivatives just mentioned work as conditions, together with another three ones, to be satisfied by the renormalized $1$PI vertex parts renormalized by multiplicative renormalization, thus connecting the bare and the renormalized correlation functions. These conditions are 
\begin{eqnarray}\label{ygfdxzsze}
\Gamma_{R}^{(2)}(0;m,g) = m^{2},
\end{eqnarray}
\begin{eqnarray}
\frac{\partial}{\partial P^{2}}\Gamma_{R}^{(2)}(0;m_{B}^{2},g)\Biggr|_{P^{2} = 0} = 1,
\end{eqnarray}
\begin{eqnarray}
\Gamma_{R}^{(4)}(P_{i}=0;m_{B},g) = g,
\end{eqnarray}
\begin{eqnarray}\label{jijhygtfrd}
\Gamma_{R}^{(2,1)}(P_{i}=0,Q_{j}=0,m_{B},g) = 1,
\end{eqnarray}
where
\begin{eqnarray}
A_{1} = \frac{(N + 8)}{6}~\parbox{10mm}{\includegraphics[scale=1.0]{fig10.eps}}_{SP}
\end{eqnarray}
\begin{eqnarray}
A_{2}^{(1)} = \frac{(N^{2} + 6N + 20)}{36}~\parbox{16mm}{\includegraphics[scale=1.0]{fig11.eps}}_{SP},
\end{eqnarray}
\begin{eqnarray}
A_{2}^{(2)} = \frac{(5N + 22)}{9}~\parbox{14mm}{\includegraphics[scale=1.0]{fig21.eps}}_{SP},
\end{eqnarray}
\begin{eqnarray}
B_{2} = \frac{(N + 2)}{18}~\parbox{12mm}{\includegraphics[scale=1.0]{fig6.eps}}^{\prime},
\end{eqnarray}
\begin{eqnarray}
B_{3} = \frac{(N + 2)(N + 8)}{108}~\parbox{12mm}{\includegraphics[scale=1.0]{fig7.eps}}^{\prime},
\end{eqnarray}
\begin{eqnarray}
C_{1} = \frac{(N + 2)}{6}\hspace{1mm}\parbox{14mm}{\includegraphics[scale=1.0]{fig14.eps}}_{SP}, 
\end{eqnarray}
\begin{eqnarray}
C_{2}^{(1)} = \frac{(N + 2)^{2}}{36}\hspace{1mm}\parbox{12mm}{\includegraphics[scale=1.0]{fig16.eps}}_{SP},
\end{eqnarray}
\begin{eqnarray}
C_{2}^{(2)} = \frac{(N + 2)}{6}\hspace{1mm}\parbox{13mm}{\includegraphics[scale=1.0]{fig17.eps}}_{SP}.
\end{eqnarray}
If we use the diagrams evaluated in \ref{Feynman integrals}, we obtain for the $\beta$-function and anomalous dimensions
\begin{eqnarray}\label{jghyuahuahuahude}
\beta(u) = u\left[ -\epsilon + \frac{N + 8}{6}\left( 1 - \frac{1}{2}\epsilon \right)\Pi u - \frac{3N + 14}{12}\Pi^{2}u^{2}\right],
\end{eqnarray}
\begin{eqnarray}\label{jghyuahuahuahudfrte}
\gamma_{\phi}(u) = \frac{N + 2}{72}\left[ \left( 1 - \frac{1}{4}\epsilon + I\epsilon \right)\Pi^{2}u^{2} - \frac{N + 8}{6}(I + 1)\Pi^{3}u^{3} \right],
\end{eqnarray}
\begin{eqnarray}\label{ujjhhahuahuahuaart}
\overline{\gamma}_{\phi^{2}}(u) = \frac{N + 2}{6}\left[\left( 1 - \frac{1}{2}\epsilon\right)\Pi u - \frac{1}{2}\Pi^{2} u^{2} \right].
\end{eqnarray}
We emphasize that the $\beta$-function and anomalous dimensions aforementioned are dependent of the symmetry point chosen through the numerical factors $-1/2$ in Eq. (\ref{jghyuahuahuahude}, $-1/4$ and $I$ (we do not need to compute explicitly the integral $I$, since it is a number and disappears in the final calculations) in Eq. (\ref{jghyuahuahuahudfrte} and $-1/2$ in Eq. (\ref{ujjhhahuahuahuaart}), respectively, in opposition to the ones calculated in the Unconventional minimal subtraction scheme where the momentum-dependent integrals cancelled out. Now evaluating the anomalous dimensions in Eqs. (\ref{jghyuahuahuahudfrte}) and (\ref{ujjhhahuahuahuaart}) at the corresponding nontrivial fixed point arising from the nontrivial solution to Eq. (\ref{jghyuahuahuahude}), after the combination of the numerical factors and the cancelling of the integral $I$ (see \ref{Feynman integrals}), which is a result of the independence of the critical exponents on the symmetry point employed in the renormalization program approached, we obtain the same LI critical exponents as the ones obtained through the earlier method, thus showing their universal character. Now we proceed to compute the all-loop quantum corrections to the critical exponents.

\section{All-loop quantum corrections to the critical exponents}\label{All-loop quantum corrections to the critical exponents}

Generalizing the expressions for the $\beta$-function and anomalous dimensions in the Unconventional minimal subtractions scheme (similar arguments can be developed in the Callan-Symanzik one as well) for any loop level \cite{Phys.Rev.D.69.105009}, we can write 
\begin{eqnarray}\label{uhgufhduhufdhsdu}
\beta(u) = u\left( -\epsilon + \sum_{n=2}^{\infty}\beta_{n}^{(0)}\Pi^{n-1}u^{n-1}\right), 
\end{eqnarray}
\begin{eqnarray}
\gamma(u) = \sum_{n=2}^{\infty}\gamma_{n}^{(0)}\Pi^{n}u^{n},
\end{eqnarray}
\begin{eqnarray}\label{uhgufhduhufdhuasd}
\overline{\gamma}_{\phi^{2}}(u) = \sum_{n=1}^{\infty}\overline{\gamma}_{\phi^{2}, n}^{(0)}\Pi^{n}u^{n}.
\end{eqnarray}
where $\beta_{n}^{(0)}$, $\gamma_{n}^{(0)}$ and $\gamma_{\phi^{2},n}^{(0)}$ are the nth-loop loop quantum corrections to the corresponding functions. We can write these expressions valid for all-loop order because the expressions for the $\beta$-functions and anomalous dimensions at next-to-leading order suggest that there is a relation between the LV and LI dimensionless renormalized coupling constants at any loop level. In fact, such relation and was shown earlier \cite{PhysRevD.84.065030, Carvalho2013850, Carvalho2014320} to be $u^{*(0)} = \Pi u^{*}$. Then, for all-loop order, the LV theory (which depends on the LV parameter $K$ only through the LV $\Pi$ factor), can be obtained from the LI one from this relation. Thus, computing the nontrivial fixed point, we get that, for any loop level, $u^{*} = u^{*(0)}/\Pi$. According to the power-law functional form of the $\beta$-functions and anomalous dimensions, we have that the LV $\Pi$ factor disappears and that the all-loop order LV critical exponents are the same as the all-loop level LI ones. This general result can be understood if we realize that the symmetry broken here is a spacetime one. It occurs in the spacetime where the field is defined and not in the internal symmetry spacetime of the field. This general result also showed that the effect of the LV composite operator operator on the values of the critical exponents had to be checked up to the end of the renormalization program and not by an initially naive analysis from power-counting considerations. 

\section{Conclusions}\label{Conclusions}

\par In this paper, we performed an analytical evaluation of the loop quantum corrections, firstly explicitly at next-to-leading level and later at all-loop order by an induction process, to the critical exponents for massive O($N$) $\lambda\phi^{4}$ scalar field theories with Lorentz violation by applying both the recently proposed Unconventional minimal subtraction and the Callan-Symanzik methods. The results for the critical exponents in both methods were identical and equal to their LI counterparts, thus conforming the universality hypothesis. The physical interpretation of this result is that the broken symmetry mechanism does not occur in the field internal symmetry spacetime but otherwise in the spacetime where the field is defined. Furthermore, this work was the first check of consistency of the Unconventional minimal subtraction scheme beyond the LI scenario, for which it was firstly proposed, for our knowledge.         

\section*{Acknowledgements}

\par GSS and PRSC would like to thank CAPES (Brazilian funding agencies) and Universidade Federal do Piau\'{i} for financial support, respectively.

\appendix
\section{Integral formulas in $d$-dimensional Euclidean momentum space}\label{Integral formulas in $d$-dimensional Euclidean momentum space}

\par Considering $\hat{S}_{d} \equiv S_{d}/(2\pi)^{d} = [2^{d-1}\pi^{d/2}\Gamma(d/2)]^{-1}$ where $S_{d} = 2\pi^{d/2}/\Gamma(d/2)$ is the unit $d$-dimensional sphere area, we have

\begin{eqnarray}
&&\int \frac{d^{d}q}{(2\pi)^{d}} \frac{1}{(q^{2} + 2pq + M^{2})^{\alpha}} = \hat{S}_{d}\frac{1}{2}\frac{\Gamma(d/2)}{\Gamma(\alpha)}\frac{\Gamma(\alpha - d/2)}{(M^{2} - p^{2})^{\alpha - d/2}},
\end{eqnarray}

\begin{eqnarray}
&&\int \frac{d^{d}q}{(2\pi)^{d}} \frac{q^{\mu}}{(q^{2} + 2pq + M^{2})^{\alpha}} = -\hat{S}_{d}\frac{1}{2}\frac{\Gamma(d/2)}{\Gamma(\alpha)}\frac{p^{\mu}\Gamma(\alpha - d/2)}{(M^{2} - p^{2})^{\alpha - d/2}},
\end{eqnarray}

\begin{eqnarray}
&&\int \frac{d^{d}q}{(2\pi)^{d}} \frac{q^{\mu}q^{\nu}}{(q^{2} + 2pq + M^{2})^{\alpha}} = \hat{S}_{d}\frac{1}{2}\frac{\Gamma(d/2)}{\Gamma(\alpha)} \nonumber \\&&   \times \left[ \frac{1}{2}\delta^{\mu\nu}\frac{\Gamma(\alpha - 1 - d/2)}{(M^{2} - p^{2})^{\alpha - 1 - d/2}} + p^{\mu}p^{\nu}\frac{\Gamma(\alpha - d/2)}{(M^{2} - p^{2})^{\alpha - d/2}} \right],
\end{eqnarray}

\begin{eqnarray}
&&\int \frac{d^{d}q}{(2\pi)^{d}} \frac{q^{\mu}q^{\nu}q^{\rho}}{(q^{2} + 2pq + M^{2})^{\alpha}} = - \hat{S}_{d}\frac{1}{2}\frac{\Gamma(d/2)}{\Gamma(\alpha)}  \nonumber \\&&   \times\left[\frac{1}{2}[\delta^{\mu\nu}p^{\rho} + \delta^{\mu\rho}p^{\nu} + \delta^{\nu\rho}p^{\mu}]\frac{\Gamma(\alpha - 1 - d/2)}{(M^{2} - p^{2})^{\alpha - 1 - d/2}} + p^{\mu}p^{\nu}p^{\rho}\frac{\Gamma(\alpha - d/2)}{(M^{2} - p^{2})^{\alpha - d/2}} \right],\nonumber \\
\end{eqnarray}

\begin{eqnarray}
&&\int \frac{d^{d}q}{(2\pi)^{d}} \frac{q^{\mu}q^{\nu}q^{\rho}q^{\sigma}}{(q^{2} + 2pq + M^{2})^{\alpha}} = \hat{S}_{d}\frac{1}{2}\frac{\Gamma(d/2)}{\Gamma(\alpha)}  \nonumber \\&&  \times\left[\frac{1}{4}[\delta^{\mu\nu}\delta^{\rho\sigma} + \delta^{\mu\rho}\delta^{\nu\sigma} +\delta^{\mu\sigma}\delta^{\nu\rho}]\frac{\Gamma(\alpha - 2 - d/2)}{(M^{2} - p^{2})^{\alpha - 2 - d/2}}  \right.  \nonumber \\  &&\left. + \frac{1}{2}[\delta^{\mu\nu}p^{\rho}p^{\sigma} + \delta^{\mu\rho}p^{\nu}p^{\sigma} + \delta^{\mu\sigma}p^{\nu}p^{\rho} + \delta^{\nu\rho}p^{\mu}p^{\sigma} +\delta^{\nu\sigma}p^{\mu}p^{\rho} +\delta^{\rho\sigma}p^{\mu}p^{\nu}]\right.  \nonumber \\  &&\left.\times\frac{\Gamma(\alpha - 1 - d/2)}{(M^{2} - p^{2})^{\alpha - 1 - d/2}}    + p^{\mu}p^{\nu}p^{\rho}p^{\sigma}\frac{\Gamma(\alpha - d/2)}{(M^{2} - p^{2})^{\alpha - d/2}} \right].
\end{eqnarray}

\section{Feynman integrals}\label{Feynman integrals}

\par By expanding the propagator in the small parameter $K_{\mu\nu}$
\begin{eqnarray}\label{expansion}
&&\frac{1}{(q^{2} + K_{\mu\nu}q^{\mu}q^{\nu} + m^{2})^{n}} = \frac{1}{(q^{2} + m^{2})^{n}}\times \nonumber \\ && \left[ 1 - n\frac{K_{\mu\nu}q^{\mu}q^{\nu}}{q^{2} + m^{2}} +  \frac{n(n+1)}{2!}\frac{K_{\mu\nu}K_{\rho\sigma}q^{\mu}q^{\nu}q^{\rho}q^{\sigma}}{(q^{2} + m^{2})^{2}} + ...\right]
\end{eqnarray}
and by applying the formulas of the \ref{Integral formulas in $d$-dimensional Euclidean momentum space} in $d = 4 - \epsilon$ we get the results to the Feynman diagrams, for their use in Sec. \ref{Unconventional minimal subtraction scheme},
\begin{eqnarray}
&&\parbox{10mm}{\includegraphics[scale=1.0]{fig10.eps}} = \frac{1}{\epsilon}\left\{ \left[1 - \frac{1}{2}\epsilon - \frac{1}{2}\epsilon L(P, m_{B}^{2}) \right]\Pi -\frac{1}{2}\epsilon K_{\mu\nu}L^{\mu\nu}(P, m_{B}^{2}) \right.  \nonumber \\  &&\left. + \frac{1}{4}\epsilon K_{\mu\nu}K_{\rho\sigma}[L^{\mu\nu}(P, m_{B}^{2})\delta^{\rho\sigma} + L^{\mu\nu\rho\sigma}(P, m_{B}^{2})] \right\},
\end{eqnarray}   

\begin{eqnarray}
&&\parbox{12mm}{\includegraphics[scale=1.0]{fig6.eps}} = \left\{-\frac{3 m_{B}^{2}}{2 \epsilon^{2}}\Bigl[1 + \frac{1}{2}\epsilon + \Bigl(\frac{\pi^{2}}{12} +1 \Bigr)\epsilon^{2} \Bigr] - \frac{3 m_{B}^{2}}{4}\tilde{i}(P, m_{B}^{2}) \right.  \nonumber \\  &&\left. -\frac{P^{2}}{8 \epsilon}\Bigl[1 + \frac{1}{4}\epsilon - 2 \epsilon L_{3}(P,m_{B}^{2})\Bigr]\right\}\Pi^{2} - \frac{3 m_{B}^{2}}{4}K_{\mu\nu}\tilde{i}^{\mu\nu}(P, m_{B}^{2}) + \nonumber \\  && \frac{P^{2}}{4}K_{\mu\nu}L_{3}^{\mu\nu}(P,m_{B}^{2}), 
\end{eqnarray}
\begin{eqnarray}
&&\parbox{12mm}{\includegraphics[scale=1.0]{fig7.eps}} = \left\{-\frac{5 m_{B}^{2}}{3 \epsilon^{3}}\Bigl[1 + \epsilon + \Bigl(\frac{\pi^{2}}{24} + \frac{15}{4} \Bigr)\epsilon^{2} \Bigr] -\frac{5 m_{B}^{2}}{2 \epsilon}\tilde{i}(P, m_{B}^{2}) \right.  \nonumber \\  &&\left. -\frac{P^{2}}{6 \epsilon^{2}}\Bigl[1+ \frac{1}{2}\epsilon - 3 \epsilon L_{3}(P,m_{B}^{2})\Bigr]\right\}\Pi^{3} - \frac{5 m_{B}^{2}}{2 \epsilon}K_{\mu\nu}\tilde{i}^{\mu\nu}(P, m_{B}^{2}) + \nonumber \\  && \frac{P^{2}}{2}K_{\mu\nu}L_{3}^{\mu\nu}(P,m_{B}^{2}), 
\end{eqnarray}
\begin{eqnarray}
&&\parbox{12mm}{\includegraphics[scale=0.8]{fig21.eps}} = \frac{1}{2\epsilon^{2}}\left\{ \left[1 - \frac{1}{2}\epsilon - \epsilon L(P, m_{B}^{2}) \right]\Pi^{2}  -\epsilon K_{\mu\nu}L^{\mu\nu}(P, m_{B}^{2}) \right.  \nonumber \\  &&\left. + \frac{1}{2}\epsilon K_{\mu\nu}K_{\rho\sigma}[L^{\mu\nu}(P, m_{B}^{2})\delta^{\rho\sigma} + L^{\mu\nu\rho\sigma}(P, m_{B}^{2})] \right\},
\end{eqnarray}  
where
\begin{eqnarray}
&& L(P, m_{B}^{2}) = \int_{0}^{1}dx\ln[x(1-x)P^{2} + m_{B}^{2}],
\end{eqnarray}
\begin{eqnarray}
&& L^{\mu\nu}(P, m_{B}^{2}) = \int_{0}^{1}dx\frac{x(1-x)P^{\mu}P^{\nu}}{x(1-x)P^{2} + m_{B}^{2}},
\end{eqnarray}
\begin{eqnarray}
&& L^{\mu\nu\rho\sigma}(P, m_{B}^{2}) = \int_{0}^{1}dx\frac{x^{2}(1-x)^{2}P^{\mu}P^{\nu}P^{\rho}P^{\sigma}}{[x(1-x)P^{2} + m_{B}^{2}]^{2}},
\end{eqnarray}
\begin{eqnarray}
&& L_{3}(P, m_{B}^{2}) = \int_{0}^{1}dx(1-x)\ln[x(1-x)P^{2} + m_{B}^{2}],
\end{eqnarray}
\begin{eqnarray}
&& L_{3}^{\mu\nu}(P, m_{B}^{2}) = \int_{0}^{1}dx\frac{x(1-x)^{2}P^{\mu}P^{\nu}}{x(1-x)P^{2} + m_{B}^{2}},
\end{eqnarray}
\begin{eqnarray}
&&\tilde{i}(P, m_{B}^{2}) = \nonumber \\ && \int_{0}^{1} dx \int_{0}^{1}dy lny \frac{d}{dy}\Bigl((1-y)ln\Bigl\{y(1-y)P^{2} + \Bigl[1-y + \frac{y}{x(1-x)}\Bigr]m_{B}^{2} \Bigr\}\Bigr),\nonumber \\
\end{eqnarray}
\begin{eqnarray}
&&\tilde{i}^{\mu\nu}(P, m_{B}^{2}) = \int_{0}^{1} dx \int_{0}^{1}dy lny \frac{d}{dy}\left\{\frac{y(1-y)^{2}P^{\mu}P^{\nu}}{y(1-y)P^{2} + \Bigl[1-y + \frac{y}{x(1-x)}\Bigr]m_{B}^{2}}\right\}.\nonumber \\
\end{eqnarray}
The diagrams $\parbox{5mm}{\includegraphics[scale=.5]{fig10.eps}}$ and $\parbox{7mm}{\includegraphics[scale=.5]{fig21.eps}}$ are evaluated perturbatively in the small parameters $K_{\mu\nu}$ up
to $\mathcal{O}(K^{2})$ and $\parbox{6mm}{\includegraphics[scale=.5]{fig6.eps}}$ and $\parbox{6mm}{\includegraphics[scale=.5]{fig7.eps}}$ up to $\mathcal{O}(K)$. The computation of the latter couple of diagrams up to $\mathcal{O}(K^{2})$ would be very tedious \cite{Carvalho2014320}. The diagrams $\parbox{6mm}{\includegraphics[scale=.5]{fig6.eps}}$ and $\parbox{6mm}{\includegraphics[scale=.5]{fig7.eps}}$ are harder to compute than the former two, because they are plagued by the problem of overlapping divergences, where the divergences proportional to $P^{2}$ and $m^{2}$, respectively, overlap. This problem can be solved by separating these divergences by using the ``partial-q'' technique \cite{THOOFT1972189}. Thus, we have

\begin{eqnarray}\label{sunset_2}
\parbox{10mm}{\includegraphics[scale=1.0]{fig6.eps}} \quad = -\frac{1}{d-3}[3m_{B}^{2}A(P, m_{B}^{2}) + B(P, m_{B}^{2})]
\end{eqnarray}
where 
\begin{eqnarray}
A(P, m_{B}^{2}) = \int \frac{d^{d}q_{1}}{(2\pi)^{d}}\frac{d^{d}q_{2}}{(2\pi)^{d}}\frac{1}{q_{1}^2 + K_{\mu\nu}q_{1}^{\mu}q_{1}^{\nu} + m_{B}^{2}}\frac{1}{q_{2}^2 + K_{\mu\nu}q_{2}^{\mu}q_{2}^{\nu} + m_{B}^{2}}\times \nonumber \\ \frac{1}{[(q_{1} + q_{2} + P)^2 + K_{\mu\nu}(q_{1} + q_{2} + P)^{\mu}(q_{1} + q_{2} + P)^{\nu} + m_{B}^{2}]^{2}},
\end{eqnarray}
\begin{eqnarray}
B(P, m_{B}^{2}) = \int \frac{d^{d}q_{1}}{(2\pi)^{d}}\frac{d^{d}q_{2}}{(2\pi)^{d}}\frac{1}{q_{1}^2 + K_{\mu\nu}q_{1}^{\mu}q_{1}^{\nu} + m_{B}^{2}}\frac{1}{q_{2}^2 + K_{\mu\nu}q_{2}^{\mu}q_{2}^{\nu} + m_{B}^{2}}\times \nonumber \\ \frac{P(q_{1} + q_{2} + P) + K_{\mu\nu}P^{\mu}(q_{1} + q_{2} + P)^{\nu}}{[(q_{1} + q_{2} + P)^2 + K_{\mu\nu}(q_{1} + q_{2} + P)^{\mu}(q_{1} + q_{2} + P)^{\nu} + m_{B}^{2}]^{2}},
\end{eqnarray}
and
\begin{eqnarray}\label{sunset_2}
\parbox{10mm}{\includegraphics[scale=1.0]{fig7.eps}} \quad = -\frac{2}{3d-10}[5m_{B}^{2}C(P, m_{B}^{2}) + 2D(P, m_{B}^{2})]
\end{eqnarray}
where 
\begin{eqnarray}
&& C(P, m_{B}^{2}) = \int \frac{d^{d}q_{1}}{(2\pi)^{d}}\frac{d^{d}q_{2}}{(2\pi)^{d}}\frac{d^{d}q_{3}}{(2\pi)^{d}}\times \nonumber \\ &&\frac{1}{q_{1}^2 + K_{\mu\nu}q_{1}^{\mu}q_{1}^{\nu} + m_{B}^{2}}\frac{1}{q_{2}^2 + K_{\mu\nu}q_{2}^{\mu}q_{2}^{\nu} + m_{B}^{2}}\frac{1}{q_{3}^2 + K_{\mu\nu}q_{3}^{\mu}q_{3}^{\nu} + m_{B}^{2}} \times \nonumber \\ && \frac{1}{(q_{1} + q_{2} + P)^2 + K_{\mu\nu}(q_{1} + q_{2} + P)^{\mu}(q_{1} + q_{2} + P)^{\nu} + m_{B}^{2}}\times \nonumber \\ && \frac{1}{[(q_{1} + q_{3} + P)^2 + K_{\mu\nu}(q_{1} + q_{3} + P)^{\mu}(q_{1} + q_{3} + P)^{\nu} + m_{B}^{2}]^{2}},
\end{eqnarray}
\begin{eqnarray}
&& D(P, m_{B}^{2}) = \int \frac{d^{d}q_{1}}{(2\pi)^{d}}\frac{d^{d}q_{2}}{(2\pi)^{d}}\frac{d^{d}q_{3}}{(2\pi)^{d}}\times \nonumber \\ &&\frac{1}{q_{1}^2 + K_{\mu\nu}q_{1}^{\mu}q_{1}^{\nu} + m_{B}^{2}}\frac{1}{q_{2}^2 + K_{\mu\nu}q_{2}^{\mu}q_{2}^{\nu} + m_{B}^{2}}\frac{1}{q_{3}^2 + K_{\mu\nu}q_{3}^{\mu}q_{3}^{\nu} + m_{B}^{2}} \times \nonumber \\ && \frac{1}{(q_{1} + q_{2} + P)^2 + K_{\mu\nu}(q_{1} + q_{2} + P)^{\mu}(q_{1} + q_{2} + P)^{\nu} + m_{B}^{2}}\times \nonumber \\ && \frac{P(q_{1} + q_{3} + P) + K_{\mu\nu}P^{\mu}(q_{1} + q_{3} + P)^{\nu}}{[(q_{1} + q_{3} + P)^2 + K_{\mu\nu}(q_{1} + q_{3} + P)^{\mu}(q_{1} + q_{3} + P)^{\nu} + m_{B}^{2}]^{2}}.
\end{eqnarray}
For the diagrams used in Sec. \ref{Callan-Symanzik method}, we have

\begin{eqnarray}
&&\parbox{10mm}{\includegraphics[scale=1.0]{fig10.eps}}_{SP} = \frac{1}{\epsilon}\left(1 - \frac{1}{2}\epsilon \right)\Pi,
\end{eqnarray}   
\begin{eqnarray}
&&\parbox{12mm}{\includegraphics[scale=1.0]{fig6.eps}}^{\prime} = -\frac{1}{8\epsilon}\left( 1 - \frac{1}{4}\epsilon +I\epsilon \right)\Pi^{2},
\end{eqnarray}  
\begin{eqnarray}
&&\parbox{12mm}{\includegraphics[scale=0.9]{fig7.eps}}^{\prime} = -\frac{1}{6\epsilon^{2}}\left( 1 - \frac{1}{4}\epsilon +\frac{3}{2}I\epsilon \right)\Pi^{3},
\end{eqnarray}  
\begin{eqnarray}
&&\parbox{12mm}{\includegraphics[scale=0.8]{fig21.eps}}_{SP} = \frac{1}{2\epsilon^{2}}\left(1 - \frac{1}{2}\epsilon \right)\Pi^{2},
\end{eqnarray}  
where the integral $I$ \cite{PhysRevD.8.4340,Amit,Carvalho2009178,Carvalho2010151}
\begin{eqnarray}
&& I = \int_{0}^{1} \left\{ \frac{1}{1 - x(1 - x)} + \frac{x(1 - x)}{[1 - x(1 - x)]^{2}}\right\}
\end{eqnarray}
is a residual number and is a consequence of the symmetry point chosen.




\end{document}